**Half-Metallic Antiferromagnet as a Prospective Material for Spintronics**

By *Xiao Hu**

*((Optional Dedication))*

[*] Xiao Hu
WPI Center for Materials Nanoarchitectonics
National Institute for Materials Science
Tsukuba 305-0044, Japan
E-mail: Hu.Xiao@nims.go.jp

Spintronics is expected as the next-generation technology based on the novel notch of spin degree of freedom of electrons. Half metals, a class of materials which behave as a metal in one spin direction and an insulator in the opposite spin direction, are ideal for spintronic applications. Half metallic antiferromagnets as a subclass of half metals are characterized further by totally compensated spin moments in a unit cell, and have the advantage of being able to generate fully spin-polarized current while exhibiting zero macroscopic magnetization. Considerable efforts have been devoted to the search for this novel material, from which we may get useful hints for prospective material exploration.

## 1. Introduction: Spintronics and Half Metal

Spintronics is a term coined for the spin-controlled electronics,[1] which is expected as a next-generation technology after the silicon-based one. The main concept of this new technology is to use the novel notch provided by the spin degree of freedom of electrons. Spintronics has the potential advantage of nonvolatility, speeding up of data processing, high circuit integration density, reduction in power consumption.

The discovery of the giant magnetoresistance (GMR) in a multilayered structure[2,3] is considered the beginning of the spintronics. As the basic block, the system consists of two magnetic layers and a nonmagnetic buffer layer in between. The directions of magnetization in the two magnetic layers can be switched between parallel to anti-parallel configurations by an external magnetic field, which results in a difference in the resistance by a factor of two.[2] It is the first control on the transport of electrons by tuning the spin degree of freedom: electrons passed through the first magnetic layer are partially spin polarized, which then pass more easily the second magnetic layer if the magnetization of the second magnetic layer is in the same direction. This scientific innovation has triggered a revolution in technology and industry, and devices such as spin valve[4] and magnetic tunnel junction[5,6] have been invented and implemented into drivers of hard discs in commercial-use personal computers.

A nominally infinite magnetorsistance can be achieved if the two magnetic layers are replaced by the so-called half metals (HMs),[7,8] a class of materials behave metallically in one spin channel due to finite density of states at Fermi level, but as an insulator (or semiconductor) in the opposite spin channel due to a gap, as schematically shown in Figure 1. HM is also unique magnetically, namely it exhibits a vanishing susceptibility at sufficiently low temperatures since magnon excitations are prohibited by the absence of states in one spin channel. For a stoichiometric system the spin magnetization per unit cell should be integer in units of Bohr magneton.

## 2. Half Metallic Antiferromagnets

Antiferromagnetism (AFM) seems to exclude spin polarization of the conduction electrons. In a conventional AFM the vanishing of the net magnetic moment originates from a symmetry relation between sites of opposite spin or the occurrence of a spin-density wave. Either effect causes the electronic structure of the two spin directions to be identical, as a result, no polarization of the conduction electrons. Van Leuken and de Groot[9] noticed that in a half-metallic material another mechanism for the exact cancellation of the local magnetic moments is provided by the requirement of the net moment to be integral: for a carefully selected material this integer can be zero.[10] Thus a complete spin polarization of the conduction electrons is not in contradiction with the notion of AFM. This class of materials was called as half metallic antiferromagnets (HMAFM).[10] The authors illustrated the strategy by substituting Ni in NiMnSb by the elements to the left of it in the periodic table, and suggested that CrMnSb in a semi-Heusler structure is a candidate of HMAFM. Calculations were performed for more complex system taking into account the crystal stability, and it was found that $V_7MnFe_8Sb_7In$ should be a HMAFM.

Pickett was the first to propose to search HMAFM in perovskite oxides,[11,12] noticing that the perovskite materials $AMO_3$, due to the simple crystal structure, potentially very large number of members, and the strong coupling between magnetic ordering and electronic properties, should be ideal. In order to achieve HMAFM, he focused on double perovskite systems which include two magnetic ions, and found $La_2VMnO_6$ is the most promising candidate. Park et al.[13] investigated the mixed-cation compound and proposed that the ordered double perovskite $LaAVRuO_6$ (A=XCa, Sr, and Ba) are strong candidates for HMAFM. Using LSDA and LSDA+U, they have shown that the HMAFM nature in $LaAVRuO_6$ is robust against (i) divalent ion replacement at A sites, (ii) oxygen site relaxation, (iii) the inclusion of the Coulomb correlation, and (iv) cation disorder.

Kodderitzsch et al. reported on the vacancy induced HM in the prototype Mott-insulating substances MnO and NiO of rocksalt structure.[14] Especially, $Ni_{0.97}O$ is claimed to be HMAFM. Taking out a Ni cation one removes eight 3d states and ten electrons from the compound, of which approximately eight are d electrons. The remaining two of the ten electrons are the s, p electrons that have transferred from Ni to oxygen to fill up the p band in the oxide material. Therefore, one loses electrons in the oxygen p bands which must be reflected by a shift of the Fermi energy into the p band. In addition, a spin polarization of the pre-dominantly oxygen p-like valence band has occurred. The important point here is that because of the spin polarization of the oxygen p bands and the chosen concentration of vacancies the Fermi energy crosses the valence oxygen p band in one spin channel only, which results in a HMAFM.

Akai et al. proposed to realize HMAFM by doping dilute magnetic ions into semiconductors.[15] Because of the existence of the gap in the host semiconductors, these impurity bands induced by the doped transition metal atoms sometimes also have a gap at the Fermi level. As long as the impurity ions carry a local magnetic moment, as in the case of dilute magnetic semiconductors, the system can easily become HM. They considered two 3d magnetic ions, for example Cr and Fe, with the sum of valence d electrons equal to 10. Based on theoretical considerations on the band structure, many-body effects, as well as first principles electronic structure calculations, they revealed that the band energy for AFM coupling among the Cr and Fe d electrons is larger than that of the ferromagnetic coupling. In this way, they obtained two candidates for HMAFM, $(Zn_{0.9}Cr_{0.05}Fe_{0.05})S$ and $(Zn_{0.9}V_{0.05}Co_{0.05})S$. See also related works.[16,17,18]

Nakao proposed[19] that monolayer superlattices CrS/FeS and VS/CoS can achieve HMAFM, where the two constituent magnetic ions have antialigned local moments that cancel exactly by virtue of the integer filling of one spin channel in stoichiometric HM. He found that the insertion of ZnS semiconducting layers in between the monolayer superlattice structures preserves the HMAFM. Taking the spin-orbit coupling as a perturbative effect in the 3d elements, the author concluded[20] that the magnetic moments induced by the spin-orbit interaction are too small to destroy the precise aspect of HM.

A variety of systems have also been investigated as the candidates of HMAFM, such as thiospinel systems,[21] octuple-perovskite cuprate,[22] Heusler alloy CoCr$_2$Al,[23] Mn(CrV)S$_4$ and Fe$_{0.5}$Cu$_{0.5}$(V$_{0.5}$Ti$_{1.5}$)S$_4$, Cr$_2$MnSb in full Heusler structure,[24] semi-Heusler compounds,[25] transition metal pnictides ABX$_2$,[26] where A and B are the transition metal elements and X=N, P, As, Sb and Bi, with the total valence d-electron number of the transition metal ions being ten, the recently fevered iron pnictides,[27, 28] and intermetallic ternary alloy Mn$_2$ZnCa.[29] In what follows, we illustrate in some detail two ways to achieve HMAFM, starting from a ferrimagnetic insulator[22] and an AFM metal,[28] respectively.

## 3. HMAFM: Perovskite Cuprates

Here we formulate a scheme to realize HMAFM by doping carrier into a perovskite cuprate[22] taking advantage of the effects of strongly correlated electrons.[30] The structure of the parent material Sr$_8$CaRe$_3$Cu$_4$O$_{24}$ is shown in Figure 2A(a).[31, 22] The first-principles calculations[32, 33, 34, 35, 36, 22] revealed that the material is an insulator (Figure 2B) with a ferrimagnetic order of 1μ$_B$ net magnetization per unit cell (Figure 2C(a)) where magnetic moments are mainly contributed from Cu atoms,[35,22] in good agreement with experiments.[31, 37, 38] As depicted in Figure 2C(a), the distribution of spin magnetization at Cu1, located at the center of the unit cell and surrounded by a perfect O octahedron (Figure 2A(a)), is of the typical e$_g$-orbital shape, and that at Cu2, surrounded by the O octahedron elongated in the directions perpendicular to the Cu1-Cu2 bonds due to the high valence of its second nearest neighboring Re$^{+7}$-ion (Figure 2A(a)),[37,35] is in form of the d$_{3z^2-r^2}$-type of e$_g$-orbital. The spin magnetizations of Cu1 and Cu2 are aligned anti-parallel respectively via the well-known super-exchange interaction.

There are small spin magnetizations -0.08 $\mu_B$ on the O2-sites located in between Cu1 and Cu2, anti-parallel to that of Cu2 (Figure 2C(a)) due to the charge-transfer effect,[39] i.e. charge fluctuations of the type $d^n \rightarrow d^{n+1}\mathbf{L}$ (**L** denotes a hole in the ligand anion valence band), as evidenced by the difference between the Hubbard energy splitting U=11.5eV and the input value $U_{eff}$=8.8eV for Cu2,[40,41,22] in contrast to U close to $U_{eff}$ for Cu1. This difference between Cu1 and Cu2 with respect to the charge-transfer effect is also consistent with the orbital ordering ($e_g$ versus $d_{3z2-r2}$) and the charge ordering ($Cu^{3+}$ versus $Cu^{2+}$) of the system.[35] In analogy to the AFM scattering of conduction electrons due to the local magnetic moment of the impurity in the Anderson model,[39] the spin magnetizations at O2 align themselves anti-parallel to that of Cu2.

The $e_g$ states of Cu1 with down-spin polarization induce a splitting in $t_{2g}$ orbits via the intra-atomic Hund's coupling, which reduces the energy of $p_x$ and $p_y$ states of O2 with down-spin through hybridization. At the O2 site the Hund's coupling due to the net local moment of $p_z$ states also enhances the exchange splitting of the $p_x$ and $p_y$ states.[42] As the result, the highest occupied bands are mainly made of spin-up 2p-states of O2 sites.

With the physics of strongly correlated electrons discussed above in mind, we modify the parent material by replacing Sr by Rb, which results in hole doping.[22] The two elements are in the same row of the periodic table, which is ideal for preserving the structure of the parent material due to the similar atom sizes.

We analyze using first-principles calculations the case in which one of the eight Sr atoms in a unit cell is substituted by a Rb atom, presuming that the material thus obtained is also uniform with an enlarged unit cell (Figure 2A(b)).[22] As seen in Figure 2B, the whole density of states remains the shape and all states with up spin shift upward to higher energy upon Rb-replacement and leave finite density of states at the Fermi level; the states with down spin remain unchanged. The doped material is thus a HM. The antiferromagnetic spin configuration of Cu1 and Cu2 is preserved perfectly as illustrated in Figure 2C(b). Since the top-most occupied states in the parent material are of spin-up 2p orbitals, the doped hole goes to O2 sites with fully-polarized up spin.[14] As the result, the local spin magnetizations of O2 change from -0.08 $\mu_B$ to -0.26 and -0.16 $\mu_B$ at the sites close and far from the doped Rb atom (Figure 2C(b)). This makes the magnetic moments compensate totally to zero in a unit cell. Therefore, we are led to conclude that $Sr_7RbCaRe_3Cu_4O_{24}$ should be a candidate of HMAFM derived from the effects of strongly correlated electrons.

## 4. HMAFM: Iron Pnicitides

Iron (Fe), one of the elements well explored from ancient time, had not been a central player in the modern electronic technologies. The recent discovery of superconductivity in iron-pnictide materials[43] certainly has opened a new chapter of study on Fe. So far the main focus on the iron-pnictide materials has been superconductivity, and the AFM order is to be suppressed as a competitor.[44]

It is intriguing to notice that the iron-pnictide materials can be good candidates for HMAFM. Since the ground state of these materials is metallic with AFM spin configuration, one only needs to open a gap in one of the two spin channels while keeping the AFM order with compensated magnetization. In the simplest picture, a Fe atom has six 3d electrons and shows an effective spin moment of $4\mu_B$ due to the intra-atomic Hund's coupling. A possible material manipulation can be substituting half of the Fe atoms by the Chromium (Cr) atom, noticing that Cr possesses four 3d electrons.[15] It is expected that the Cr atom will not change the AFM order of compensated magnetization of the parent material, while modify the band structure due to the different atomic number from Fe.[27,28]

To be specific, we focus on a typical iron-pnictide BaFe$_2$As$_2$.[45] In this material, Fe ions are tetrahedrally coordinated by As atoms and form the Fe-As layer, which is sandwiched by Ba layers. We concentrate on the situation that half of the Fe atoms are substituted by Cr atoms. As revealed by the first principles calculations,[33,34, 28] the ground state of BaCrFeAs$_2$ is characterized by three-dimensional intervening lattices of Fe and Cr ions, with a $\sqrt{2}\times\sqrt{2}\times1$ unit cell as displayed in Figure 3A(a). The magnetization is mainly contributed by Fe (-2.62 $\mu_B$) and Cr (2.75 $\mu_B$) ions (Figure 3A(b)), while each As atom carries a small magnetic moment (-0.08 $\mu_B$) parallel to that of Fe ions due to the hybridization between As:4p and transition metal:3d electrons. The Ba layers act as the charge reservoir with negligible magnetic moment. According to the calculations, the magnetic moments of the Fe, Cr and As atoms compensate completely, resulting in the zero total magnetization in the unit cell as expected.[28]

Let us look into the electronic band structure of the new material BaCrFeAs$_2$ (Figure 3B). In the spin-up channel, Cr nominally has four 3d electrons, and thus its majority 3d band is not fully occupied. The hybridization between As:4p and Cr:t$_{2g}$ states forms the bonding states deep in the valence band, and partially occupied anti-bonding states around the Fermi level. Although there exist Fe:3d-As:4p as well as Fe:3d-Cr:3d hybridizations, most of the Fe:3d minority states remain non-bonding and locate around the Fermi level.[28] Therefore, as far as the spin-up channel concerned, BaCrFeAs$_2$ is metallic, similar to the parent material.[46] In the spin-down channel, Fe:3d majority states lie deeply in the valence band and are fully occupied. Most of the minority states of Cr are above the Fermi level and form the bottom of conduction band, in contrast with Fe in the spin-up channel due to the difference in the electronegativity between Cr and Fe.[28]

Since the Cr:3$d_{xy}$ orbitals reside on the Fe-As plane with their petals pointing to the nearest neighboring Cr ions along (110) or (1-10) direction, and the As atoms just locate above the center of the transition-metal squares (Figure 3A(a)), the As:4$p_z$ orbitals mediate the exchange interaction between the Cr:3 $d_{xy}$ states (Figure 3C). Similarly, the As:4$p_y$ (As:4$p_x$) state provides a bridge for the exchange coupling between the two Cr:3 $d_{xz}$ (Cr:3 $d_{yz}$) states. As the result, the spin-down As:4p states in the vicinity of the Fermi level hybridize with a part of the Cr:$t_{2g}$ minority states, and lower their energies to the valence band, as evidenced by the band-decomposed charge density[33,34] depicted in Figure 3C. A gap of ~0.3 eV is thus opened in the spin-down channel (Figure 3B).[28]

Combining the totally compensated magnetization in unit cell and the HM band structure, we expect BaCrFeAs$_2$ to be a candidate of HMAFM. The above picture is also supported by a tight-binding analysis.[47,28] It is noteworthy that the basic lattice structure of the transition metal plane of the iron pnictides is very helpful for the realization of HMAFM.

## 5. Discussions

Despite of considerable efforts devoted to the search for HMAFM, there is no clear confirmation of this novel material by the time of writing. The asymmetry with respect to spin resolved density of states at the Fermi level on one hand, and the compensation between spin magnetizations in the two spin channels on the other hand, do require a fine engineering on the band structures, which makes HMAFM rare. The sophisticated band structure of HMAFM needs complex crystal structures, and there are more chances for lattice distortions to occur, which in turn would suppress the bulk HMAFM property. Spin-orbital interactions may distort the complete compensation of spin moments. Since the spin magnetizations in the two directions are contributed by different ions in different environments, their temperature dependences can be different, which may make a totally compensated spin magnetization only available at zero temperature.[25] As a technical reason, experimental methods sensitive to surface property may hardly give a real picture for the bulk property of HMAFM, which makes experimental confirmation of any HMAFM complex.

The present paper does not include a discussion on first principles calculations due to the limited space, which, as a matter of fact, are involved in almost all the theoretical searches for HMAFM. For this important part readers are recommended to refer to the original papers.

## 6. Conclusions

Half metallic antiferromagnets are not an oxymoron since the transport properties are dominated by the density of states at the Fermi level, while the spin magnetization is the accumulation over all occupied states including those of low energies. The total compensation of spin magnetizations in half metallic antiferromagnets in stoichiometric systems is protected by the spin quantization guaranteed by the gap in the density of states at the Fermi level in one spin channel. Without stray magnetic field while keeping spin sensitivity, half metallic antiferromagnets are ideal for novel spin polarized STM tips,[9] since they do not disturb the magnetic property of the material surface under detection. Half metallic antiferromagnets can work idealy as the anchor layer in a spin valve, since theiry thin films do not suffer from domain walls that occur in magnetic materials including half metals with net magnetization. Half metallic antiferromagnets can also be the parent materials for single spin superconductivity with triplet Cooper pairs due to the finite density of states in only one spin channel.[12] All these unique properties of half metallic antiferromagnets certainly will motivate continuing effort towards the realization of this class of novel functional materials, and their applications in advanced spintronics applications.

## Acknowledgements

The author is grateful for X.-G. Wan, M. Kohno, Y.-M. Nie and S.-J. Hu for collaborations. The works were supported by the WPI Initiative on Materials Nanoarchitectonics, MEXT, Japan.

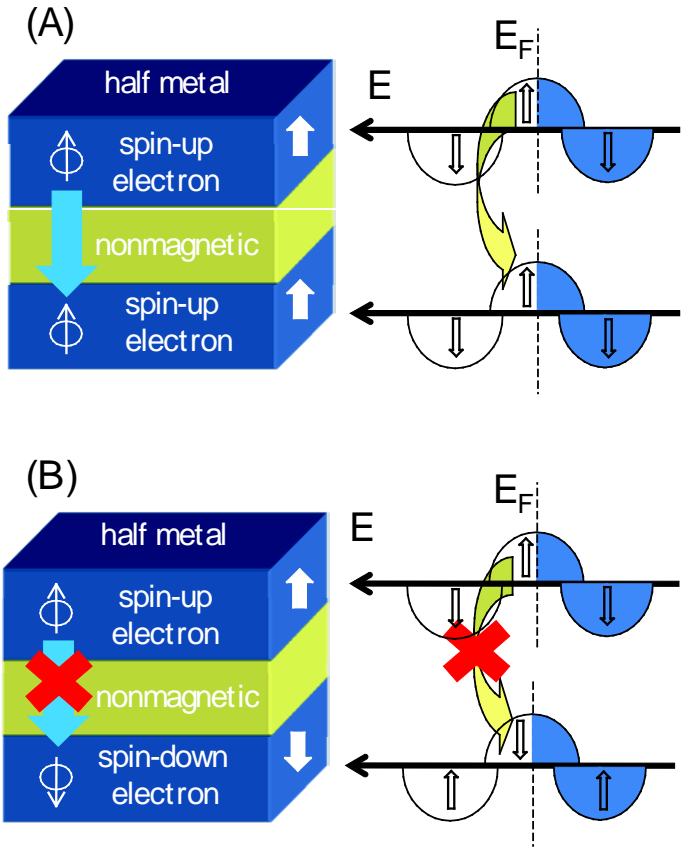

**Figure 1.** Two possible configurations for a multilayered structure of two half-metal thin films sandwiching a nonmagnetic film, which generate large difference in conductance. (A) The spin polarizations of electrons at the Fermi level in the two half-metal thin films are parallel (upward in this case), and thus a current can pass through the system. (B) Nominally no current can pass through when the spin polarizations of electrons at the Fermi level are opposite if spin scattering is not considered in the nonmagnetic layer. The bold white arrow at the side of half-metal thin films should be considered as the spin direction in which density of states is finite at the Fermi level, rather than the direction of total magnetization.

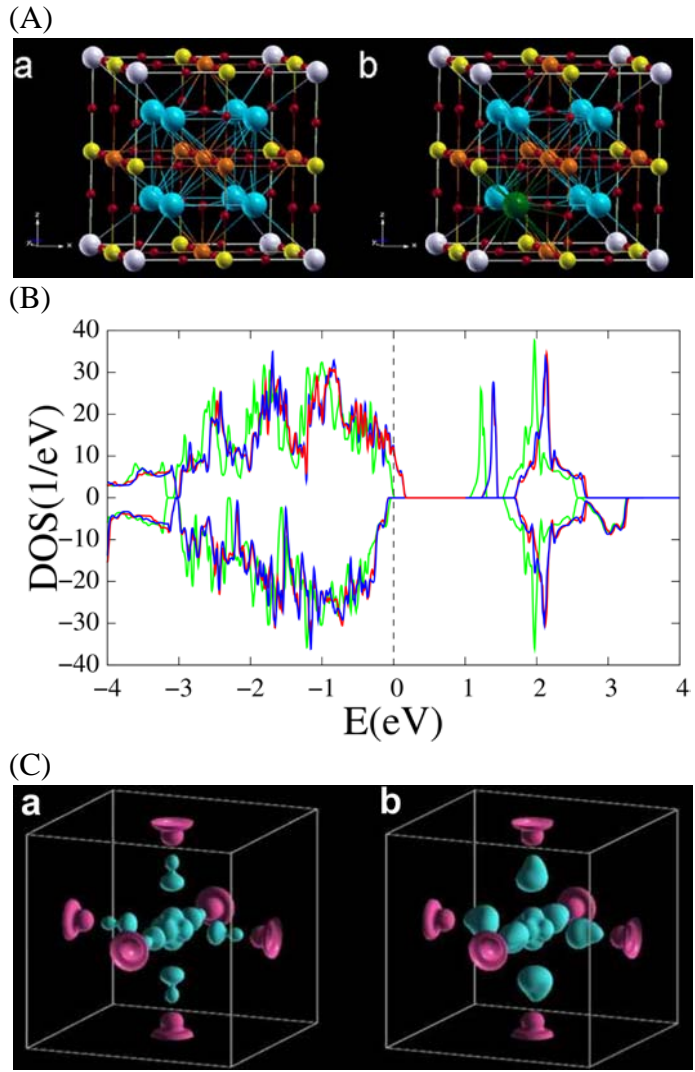

**Figure 2.** (A) (a) Unit cell for the parent material $Sr_8CaRe_3Cu_4O_{24}$ and (b) that for the modified material $Sr_7RbCaRe_3Cu_4O_{24}$ with the lowest symmetry. The Sr (blue), Rb (green), Ca (gray), Re (yellow), Cu (brown), and O (red) atoms are shown in colored spheres. (B) Spin-resolved density of states for the parent material (green), and that for the modified material of unit cell with the lowest /highest symmetry (red/blue). (C) Isosurface of spin magnetization density at $\pm 0.07\ \mu_B/A^3$ with pink/blue for up/down spin for the parent material (a) and the modified material (b). Reproduced with permission.[22] Copyright 2008, American Physical Society.

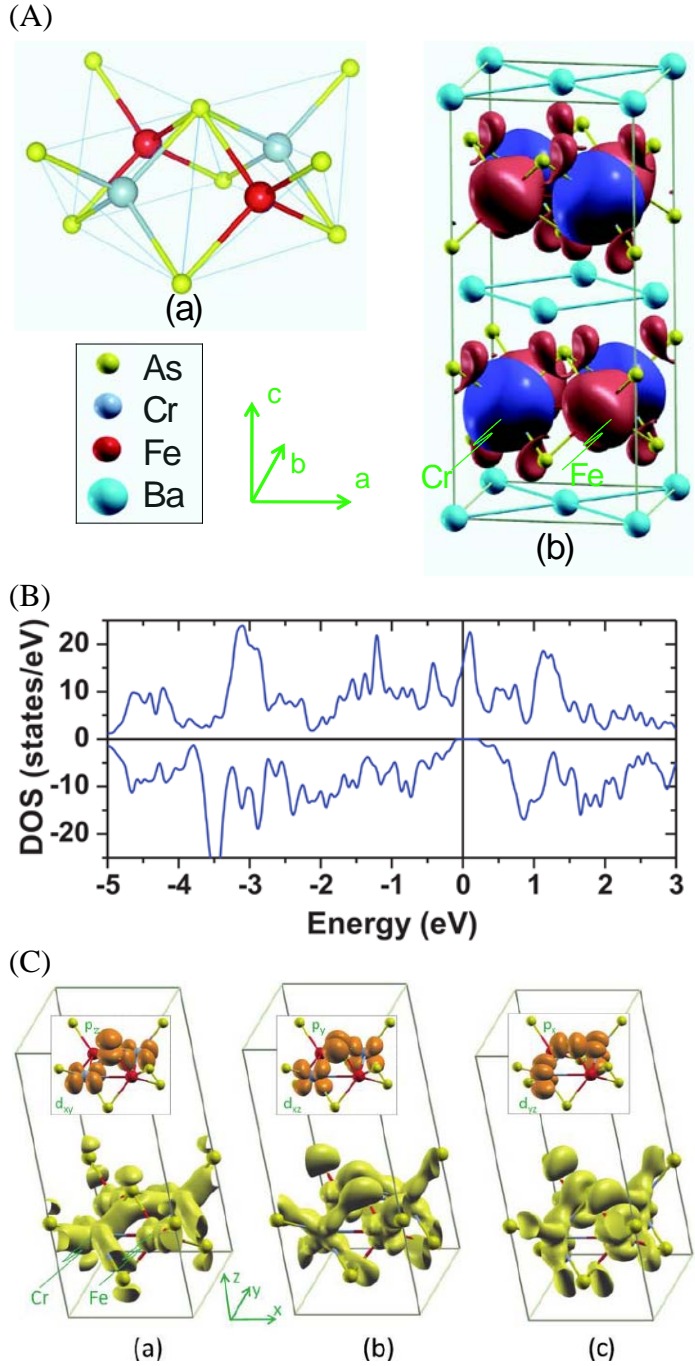

**Figure 3.** (A) (a) Crystal structure of the transition metal –As layer of $BaCrFeAs_2$, and (b) spin density of the antiferromagnetic state (blue/red surface: spin-up/down) with the isovalue $\pm 0.025$ $\mu_B/A^3$. (B) Spin-resolved density of states of $BaCrFeAs_2$. (C) Band-decomposed charge density: (a) $Cr:3d_{xy}$ bonding via $As:4p_z$, (b) $Cr:3d_{xz}$ bonding via $As:4p_y$ and (c) $Cr:3d_{yz}$ bonding via $As:4p_x$ at the $\Gamma$ point of the spin-down valence band maximum associated with the unit cell of $\sqrt{2} \times \sqrt{2} \times 1$. Only one transition metal-As layer is shown. Isovalue = 0.004 for (a) and 0.008 $e/A^3$ for (b) and (c). Insets schematically show the corresponding atomic orbitals of Cr and As. Reproduced with permission.[28] Copyright 2010, American Chemical Society.